\documentclass[twocolumn,showpacs,preprintnumbers,amsmath,amssymb]{revtex4}

\usepackage{epsf}
\usepackage{graphicx}  % Include figure files
\usepackage{dcolumn}   % Align table columns on decimal point
\usepackage{bm}        % bold math

\newcommand{\be}{\begin{equation}}
\newcommand{\en}{\end{equation}}
\newcommand{\bea}{\begin{eqnarray}}
\newcommand{\ena}{\end{eqnarray}}
\newcommand{\hbo}{\hbox to 1 true cm {\hfill } }

\begin{document}

\preprint{UNITUE-THEP/2-2004}

\title{Signals of confinement in Green functions of SU(2) Yang-Mills theory }

\author{Jochen Gattnar}
\author{Kurt Langfeld}
%\email{kurt.langfeld@uni-tuebingen.de} 
\author{ Hugo Reinhardt} 

\affiliation{
Insitut f\"ur Theoretische Physik, Universit\"at T\"ubingen\\ 
D-72076 T\"ubingen, Germany. 
}

\date{ March 9, 2004}

\begin{abstract}
The vortex picture of confinement is employed to explore the signals  
of confinement in Yang-Mills Green functions. By using SU(2) 
lattice gauge theory, it has been well established that the 
removal of the center vortices from the lattice configurations results 
in the loss of confinement. The running coupling constant, the gluon 
and the ghost form factors are studied in Landau gauge for both 
cases, the full and the vortex removed theory. In the latter case, a strong 
suppression of the  running coupling constant and the gluon 
form factor at low momenta is observed. At the same time, the singularity 
of the ghost form factor at vanishing momentum disappears. This 
observation establishes an intimate correlation between the ghost 
singularity and confinement. The result also shows that 
a removal of the vortices generates a theory for which Zwanziger's 
horizon condition for confinement is no longer satisfied. 
\end{abstract}


\maketitle

Since Quantum Chromo Dynamics (QCD) has been 
recognized as the theory of strong 
interactions, a major challenge has been to explain confinement of 
quarks from first principles. 

\vskip 0.3cm 
More than twenty years ago, Mandelstam and 't~Hooft conjectured that 
topological obstructions of the gauge field, such as monopoles 
and vortices, might play a key role for understanding 
confinement. By means of lattice simulations evidence has been 
accumulated that center vortices are responsible for 
confinement (see~\cite{Greensite:2003bk} for a recent review): The quark 
antiquark potential calculated with vortex configurations reproduces the 
linear rising potential~\cite{DelDebbio:1996mh,DelDebbio:1998uu}. 
On the other hand, a removal of the vortices from the lattice 
configurations ``by hand'' results in the loss of confinement. 
Secondly, it was observed that the properties of the confining vortices 
nicely extrapolate to the continuum limit for 
$SU(2)$~\cite{Langfeld:1997jx} and $SU(3)$~\cite{Langfeld:2003ev}. 
Furthermore, in the vortex picture the finite temperature 
deconfinement phase transition appears in SU(2) as a vortex 
depercolation transition~\cite{Langfeld:1998cz,Engelhardt:1999fd}. 
Indeed, vortex projected ensembles nicely reflect the 
correct  universality class of the transition in 
SU(2)~\cite{Langfeld:2003zi}. In addition, models based upon the 
center vortex picture nicely reproduce the order of the 
deconfinement phase transition for both 
$SU(2)$~\cite{Engelhardt:1999wr} and $SU(3)$~\cite{Engelhardt:2003wm}. 

\vskip 0.3cm 
A different confinement picture was proposed by Gribov~\cite{Gribov:1977wm}
and further elaborated by Zwanziger~\cite{Zwanziger:1991ac}. 
This picture makes use of QCD Green functions in Landau 
gauge, the IR properties of which presumably encode the information on 
confinement~\cite{Zwanziger:1991ac}. It is argued that the gauge field 
configurations which are relevant in the thermodynamic limit are 
concentrated on the Gribov horizon~\cite{Zwanziger:1991ac}. In this case, 
the ghost form factor in Landau gauge would diverge at zero momentum 
transfer showing that the above ``horizon condition'' is satisfied. 
Suman and Schilling~\cite{Suman:1995zg} obtained first indications 
using lattice simulations, that the ghost propagator in Landau gauged 
SU(2) theory is indeed more singular in the infrared than the free 
ghost propagator. 

\vskip 0.3cm 
Both pictures of confinement, the center vortex picture and the 
Gribov-Zwanziger picture, are compatible given the fact that 
center vortex configurations lie on the Gribov 
horizon~\cite{Greensite:2004ke} (The argument presented 
in~\cite{Greensite:2004ke} can be extended to Landau gauge). 

In this letter, we explicitly establish a relation between both 
confinement pictures. We will show that the removing of 
the center vortices eliminates the confinement signals from the 
Green functions: the singularity of the ghost form factor 
disappears. As a byproduct, we will also show that the running coupling 
constant is largely suppressed in the IR regime when the center vortices 
are removed.  These findings are in accordance with the results 
in~\cite{Langfeld:2001cz} where a suppression of the gluon 
form factor was found upon vortex removal. These results underline the 
importance of the vortices as IR effective degrees of freedom. 
Preliminary results obtained by one of us have been presented 
in~\cite{Langfeld:2002bg}. 

\vskip 0.3cm 
The investigation of the IR properties of Yang-Mills Green functions 
necessarily involves non-perturbative techniques. Besides lattice 
simulations, Dyson-Schwinger equations 
(DSE)~\cite{Roberts:2000aa,Alkofer:2000wg}, variational 
techniques~\cite{Szczepaniak:2001rg,Szczepaniak:2003ve,Feuchter:2004gb} 
and flow equations~\cite{Jungnickel:mn,Pawlowski:2003hq}
have been applied. In particular, Dyson-Schwinger equations have 
attracted much interest over the last decade due to their usefulness 
for the description of the physics of hadrons. 
Let us focus onto the results from the DSEs in Landau gauge. 
Let $F_R(p^2, \mu ^2)$ and $J_R(p^2,\mu ^2)$ 
denote the form factors (for a renormalization point $\mu $) 
of the gluon and the ghost propagator, respectively.  
These form factors have been recently obtained from the 
coupled set of continuum DSEs in Landau gauge by several groups
\cite{vonSmekal:1997is}--\cite{Bloch:2003yu}.
%,vonSmekal:1997is2,Atkinson:1997tu,Atkinson:1998zc,
%Zwanziger:2001kw,Bloch:2001wz,Lerche:2001thesis,Lerche:2002ep,
%Fischer:2002hn,Fischer:2002eq,Zwanziger:2002ia,Bloch:2003yu}.   
In~\cite{vonSmekal:1997is,vonSmekal:1997is2}, it was firstly pointed out 
that the gluon and ghost form factors might satisfy simple scaling laws 
in the IR momentum range 
$p \ll  1 \, \mathrm{GeV} $
\be 
F_R(p^2, \mu^2) \; \propto \left[ p^2 \right]^{\alpha} \; , \hbo 
J_R(p^2, \mu^2) \; \propto \left[ p^2 \right]^{\beta} \;
\label{eq:DSEpred}  
\en 
where the following remarkable sum rule holds for the IR exponents 
$\alpha $ and $\beta $: 
\be 
\alpha \; + \; 2 \, \beta \; = \; 0 \, . 
\label{eq:sumrule}  
\en 
It turns out that this result is rather independent of the
truncation scheme under consideration although the precise values 
for $\alpha $ and $\beta $ strongly depends on the truncation 
scheme (see, e.g., ~\cite{Lerche:2002ep,Bloch:2003yu}). 
Disregarding the Gribov ambiguity, Taylor has shown that the 
ghost-ghost-gluon-vertex renormalization constant is finite in 
Landau gauge~\cite{Taylor:ff} at least to all orders perturbation theory. 
If this comes true at the genuine non-perturbative level, the running 
coupling can be defined from the form factors by~\cite{vonSmekal:1997is}
\be
\alpha_R(p^2) \; = \; 
\alpha_R(\mu^2) \; F_R(p^2,\mu ^2) \; J_R^2(p^2, \mu^2) \; . \
\label{eq:run}  
\en
The IR sum rule (\ref{eq:sumrule}) also implies that the running
coupling develops a fixed point in the IR limit 
\be 
\lim_{p\to 0} \alpha_R(p^2) \; = \;
\alpha_c \;=\;\hbox{constant}. 
\label{eq:alphac}
\en 
Note that this result is independent of the values of $\alpha $, $ \beta $
as long as the IR sum rule (\ref{eq:sumrule}) is satisfied. 

\begin{figure}[t] 
%\centering
\centerline{ 
\epsfxsize=7cm \epsffile{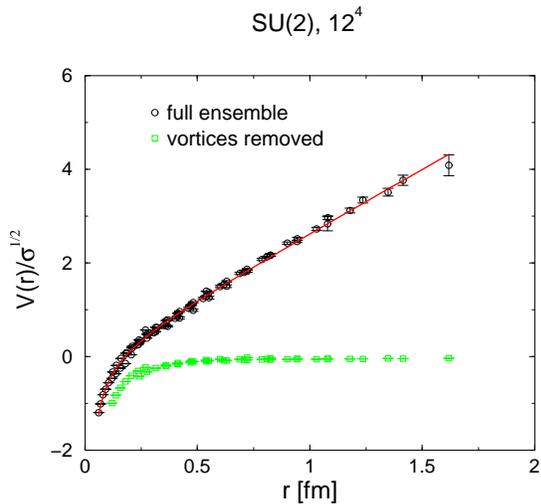}
}
\caption{ The static quark potential $V(r)$ in units of the string 
   tension $\sigma $ as function of the 
   quark antiquark distance $r$ for full SU(2) Yang-Mills theory 
   and for the case of the modified theory. Plot 
   from~\cite{Langfeld:2001ek}. 
\label{fig:1}
}
\end{figure}
\vskip 0.3cm  
Many efforts were devoted to determine the gluon and ghost form factors 
by lattice 
simulations~\cite{Suman:1995zg},~\cite{Cucchieri:1997fy}--\cite{Bakeev:2003rr}. 
%% Cucchieri:1999sz, Langfeld:2001cz, Bonnet:2001uh, Cucchieri:1997dx, 
%% Bakeev:2003rr 
A recent combined study of the gluon/ghost formfactor 
and the running coupling constant was presented in~\cite{Bloch:2003sk}. 
There, it was found for the first time using lattice simulations 
that the ghost-ghost-gluon-vertex renormalization 
constant is indeed finite. This paves the path 
to the definition of the running coupling constant (\ref{eq:run}) via 
2-point Green functions only. Moreover, the lattice data confirm the sum 
rule (\ref{eq:sumrule}) and are consistent with an IR fixed point 
value of $\alpha_c = 5(1)$ for two colors. 
However, further lattice studies are desirable 
to determine the IR scaling exponents $\alpha $ and $\beta $ in a 
reliable way.

\vskip 0.3cm
Here, we will work out 
the signature of confinement in the ghost form factor and 
the running coupling constant, and will confirm our earlier findings 
on the IR suppression of the gluon form factor when vortices are 
removed~\cite{ Langfeld:2001cz}. 
The present lattice simulations were carried out on a 
$16^3 \times 32 $ lattice using Wilson action with 
$\beta $ values ranging from $2.15$ to $2.5$. Physical units 
are obtained by eliminating the lattice spacing $a$ using the 
measured values for $\sigma a^2$. A string tension 
$\sigma = 440 \, $MeV was used as reference scale. The simulation 
parameters are listed in table \ref{tab:1}. 

\begin{table}[ht]
\begin{center}
\begin{tabular}{lcccccc} \hline
$\beta $ & $2.15$ & $2.2$ & $2.3$ & $2.375 $ & $2.45$ & 
$2.525$  \\ \hline 
$\sigma a^2$ &  $0.28(1)$ & $0.220(9) $ & $0.136(2)$ & $0.083(2)$ 
& $0.0507(8)$  & $0.0307(5)$ \\ \hline 
\end{tabular}
\end{center}
\vskip -5mm
\caption{ Lattice spacing $a$ in units of the string tension $\sigma $ 
  for the values $\beta $ used in the present simulation. }
\label{tab:1}
\vskip 5mm
\end{table} 

\vskip 0.3cm
In a first step, we identify the $Z_2$ vortex configurations for a particular 
SU(2) lattice configuration by the method of center projection after 
maximal center gauge fixing. This procedure, described in detail 
in reference~\cite{DelDebbio:1998uu}, 
generates  a vortex structure which nicely extrapolates to the 
continuum limit. Once the vortices are identified, we used the 
Forcrand D'Elia method~\cite{deForcrand:1999ms}
to remove the vortices from the lattice configurations. The quark 
antiquark potential obtained in this way is shown in figure~\ref{fig:1} 
in comparison with the full potential. Obviously, configurations 
from which the vortices were removed lack the capability to confine 
quarks. 

\begin{figure}[t] 
%\centering
\centerline{ 
\epsfxsize=7cm \epsffile{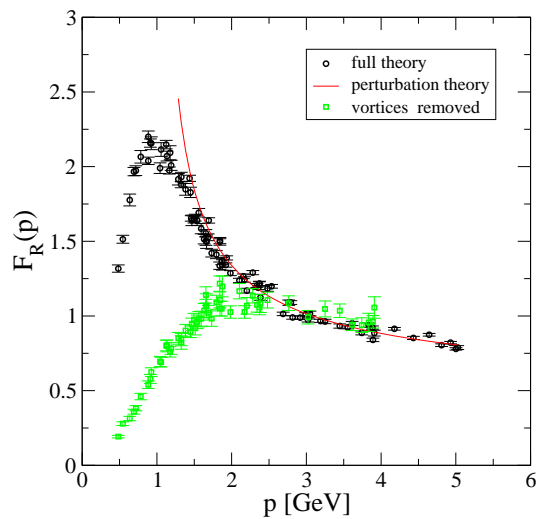}
}
\caption{ The renormalized gluon form factor as function of 
   the momentum transfer $p$ for full SU(2) gauge theory and 
   for the vortex removed theory, respectively. 
\label{fig:2}
}
\end{figure}
\vskip 0.3cm 
Subsequently, Landau gauge is implemented for both theories, 
the full SU(2) Yang-Mills theory and its vortex removed counterpart, 
and the gluon and ghost form factors are extracted from both lattice 
ensembles. Landau gauge fixing and the calculation of the form factors 
are described in detail in~\cite{Bloch:2003sk}. 
At high momenta, the momentum dependence is well described by 
\be
F_R (p^2 , \mu^2) , \; J_R (p^2 , \mu^2)
\; \approx \; d_2(\mu) \; \biggl[  \alpha _{2-loop} \Bigl(
\frac{p^2}{\Lambda ^2_{2-loop}} \Bigr) \biggr]^{\gamma} 
\label{eq:form2loop}
\en
where $\mu $ is the renormalization point, 
$\gamma$ is the leading-order anomalous dimension of the gluon 
(respectively ghost) propagator, given by $\gamma = 13/22 $ 
(respectively $\gamma = 9/44 $). The running coupling constant at 
2-loop level is independent of the renormalization scheme and for 
SU(2) given by 
\be
\alpha _{2-loop} \Bigl( x= \frac{p^2}{\Lambda ^2_{2-loop}} \Bigr) 
\;=\;
\frac{ 6 \pi }{11 \, \ln x } \Biggl\{
1 \; - \; \frac{ 102 }{ 121 } \frac{ \ln ( \ln x )
}{ \ln x } \Biggl\} 
\; . 
\label{eq:2-loop}
\en
It turns out that the choice 
$$
\Lambda _{2-loop} \; \approx \; 950 \; \mathrm{MeV} 
$$
well describes the lattice data  (see~\cite{Bloch:2003sk} for details). 

\vskip 0.3cm 
Here, we confirm our earlier findings~\cite{Langfeld:2001cz} for the gluon 
formfactor (see figure \ref{fig:2}): the gluon 
form factor of the vortex removed theory is suppressed in the intermediate 
momentum range, while the UV regime is unchanged to a large extent. 

\begin{figure}[t] 
%\centering
\centerline{ 
\epsfxsize=7cm \epsffile{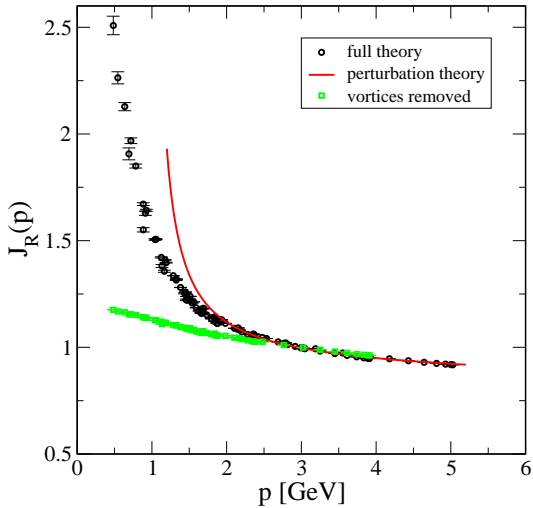}
}
\caption{ The renormalized ghost form factor as function of 
   the momentum transfer $p$ for full SU(2) gauge theory and 
   for the non-confining theory, respectively. 
\label{fig:3}
}
\end{figure}
\begin{figure}[t] 
%\centering
\centerline{ 
\epsfxsize=7cm \epsffile{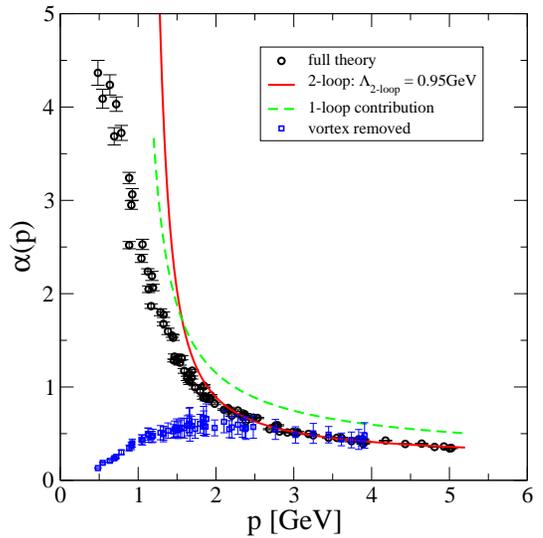}
}
\caption{ The running coupling constant (\ref{eq:run}) defined in Landau 
   gauge as function of 
   the momentum transfer $p$ for full SU(2) gauge theory and 
   for the non-confining theory, respectively. 
\label{fig:4}
}
\end{figure}

\vskip 0.3cm 
The central results of the present paper are the ghost formfactor 
and the running coupling constant 
of the vortex removed theory. Our numerical results are summarized 
in figures \ref{fig:3} and \ref{fig:4}. While the ghost form factor 
of full SU(2) gauge theory diverges in the IR limit 
(see e.g.~\cite{Suman:1995zg,Bloch:2003sk}), our results suggest that it 
approaches a constant only slightly above $1$ in the IR limit for the 
non-confining theory. Since the IR divergence is 
related to the proximity of the gauge configurations to the 
first Gribov horizon, our findings strongly support 
the Gribov--Zwanziger confinement picture. We here point out 
that Kugo and Ojima proposed a confinement criterion based upon  
the framework of the BRST quantization: this criterion signals that the 
physical subspace only consists of color singlet states~\cite{Kugo:gm}. 
In Landau gauge, this criterion is fulfilled if the 
ghost form factor is singular at zero momentum~\cite{Kugo:1995km}. 
It therefore coincides with Zwanziger's horizon condition. We stress, 
however, that for the derivation of the Kugo-Ojima criterion, one assumes 
that a BRST charge operator is uniquely defined for the whole configuration 
space within the first Gribov horizon. 
At the present stage of investigations, this assumption is 
unjustified due to the presence of Gribov ambiguities~\cite{thanks}.

\vskip 0.3cm 
Finally, let us focus onto the imprint of the confining vortices 
on the running coupling strength $\alpha _R (p)$. 
Figure \ref{fig:4} shows the running coupling 
for both, the full SU(2) theory and the one with the center 
vortices removed. 
Since the ghost formfactor approaches a constant in the vortex removed 
case and the gluon form factor is suppressed in the IR regime 
compared with the free one, it does not come as a surprise 
that the running coupling strength vanishes in the IR limit. 

\vskip 0.3cm 
In conclusion, SU(2) Yang-Mill theory looses its capability to confine 
quarks when the confining vortices 
were removed by the Forcrand D'Elia procedure~\cite{deForcrand:1999ms}. 
At the same time, the divergence of 
the ghost formfactor at vanishing momentum disappears. Our findings 
therefore establish a connection between the vortex picture 
of confinement and the Gribov-Zwanziger confinement criterion. 
We furthermore find that the strength of the running coupling 
constant is drastically reduced in the intermediate momentum 
region. This also indicates a tight relation between the 
vortex picture and the spontaneous chiral symmetry breaking. 
The latter to happen, the integrated strength at intermediate 
momenta must exceed a critical value.

\vspace{.3cm}
{\bf Acknowledgments.} 
KL acknowledges the support of this work by the Landesforschungsschwerpunkt 
{\it Quasiteilchen}, and HR by the Deutsche Forschungsgemeinschaft under 
contract DFG--Re856/5-1.

\end{document}